\documentclass[twocolumn,showpacs,aps,prl,superscriptaddress]{revtex4}
\usepackage{graphicx}
\usepackage{dcolumn}
\usepackage{epsfig}
\usepackage{amsmath}
\RequirePackage{xspace}

\newcommand{\BaBarYear}    {06}
\newcommand{\BaBarNumber}  {051}
\newcommand{\SLACPubNumber} {11992}

\input babarsym

\def\epem  {\ensuremath{e^+e^-}\xspace}
\newcommand\dbline{\noalign{\vskip 0.10truecm\hrule}\noalign{\vskip 2pt}\noalign{\hrule\vskip 0.10truecm}}
\providecommand{\tbline}{\noalign{\vskip 0.05truecm\hrule\vskip0.05truecm}}

\newcommand\etal{{\it et al.}}
\newcommand{\bma}[1]{\boldmath{$#1$}}
\newcommand{\half}{\ensuremath{{1\over2}}}
\newcommand{\pvec}{{\bf p}}
\newcommand{\calB}{\ensuremath{{\cal B}}}
\providecommand{\bfemsix}{${\cal B} (10^{-6})$}

\newcommand{\DE}{\ensuremath{\Delta E}}
\newcommand{\UfourS}{\ensuremath{\Upsilon(4S)}}
\newcommand{\thetaT}{\ensuremath{\theta_{\rm T}}}
\newcommand{\costhr}{\ensuremath{\cos\thetaT}}
\newcommand{\xf}{\ensuremath{{\cal F}}}
\newcommand{\hel}{\ensuremath{{\cal H}}}
\newcommand{\mres}{\ensuremath{m_{\rm res}}}

\newcommand{\signf}{$\cal S$ ($\sigma$)}
\newcommand{\eff}{$\epsilon$ (\%)}

\newcommand{\etagg}{\ensuremath{\eta_{\gaga}}}
\newcommand{\etappp}{\ensuremath{\eta_{3\pi}}}
\newcommand{\etatogg}{\ensuremath{\eta\ra\gaga}}
\newcommand{\etatoppp}{\ensuremath{\eta\ra\pi^+\pi^-\pi^0}}

\newcommand{\etaptoepp}{\ensuremath{\etapr\ra\eta\pip\pim}}
\newcommand{\etapepp}{\ensuremath{\etapr_{\eta\pi\pi}}}
\newcommand{\etaprg}{\ensuremath{\etapr_{\rho\gamma}}}
\newcommand{\etaptorg}{\ensuremath{\etapr\ra\rho^0\gamma}}

\newcommand{\fetaeta}{\ensuremath{\eta\eta}}
\newcommand{\etaeta}{\ensuremath{\Bz\ra\fetaeta}}
\newcommand{\retaeta}{\ensuremath{1.1^{+0.5}_{-0.4}\pm0.1}}
\newcommand{\Betaeta}{\ensuremath{\calB(\etaeta)}}

\newcommand{\uletaeta}{\ensuremath{1.8}}

\newcommand{\fetaggetagg}{\ensuremath{\eta_{\gamma\gamma}\eta_{\gamma\gamma}}}
\newcommand{\fetaggetappp}{\ensuremath{\eta_{\gamma\gamma}\eta_{3\pi}}}
\newcommand{\fetapppetappp}{\ensuremath{\eta_{3\pi}\eta_{3\pi}}}

\newcommand{\fetaphi}{\ensuremath{\eta\phi}}
\newcommand{\retaphi}{\ensuremath{0.1\pm 0.2\pm0.1}}
\newcommand{\Betaphi}{\ensuremath{\calB(\etaphi)}}
\newcommand{\etaphi}{\ensuremath{\Bz\ra\fetaphi}}

\newcommand{\uletaphi}{\ensuremath{0.6}}

\newcommand{\fetaggphi}{\ensuremath{\eta_{\gamma\gamma}\phi}}
\newcommand{\fetapppphi}{\ensuremath{\eta_{3\pi}\phi}}

\newcommand{\fetaggkz}{\ensuremath{\eta_{\gamma\gamma} K^0  }}
\newcommand{\fetapppkz}{\ensuremath{\eta_{3\pi} K^0}}

\newcommand{\fetapetap}{\ensuremath{\etapr\etapr}}
\newcommand{\etapetap}{\ensuremath{\Bz\ra\fetapetap}}
\newcommand{\Betapetap}{\ensuremath{\calB(\etapetap)}}
\newcommand{\retapetap}{\ensuremath{1.0^{+0.8}_{-0.6}\pm0.1}}

\newcommand{\uletapetap}{\ensuremath{2.4}}

\newcommand{\fetapeppetapepp}{\ensuremath{\etapr_{\eta\pi\pi}\etapr_{\eta\pi\pi}}}
\newcommand{\fetapeppetaprg}{\ensuremath{\etapr_{\eta\pi\pi}\etapr_{\rho\gamma}}}

\newcommand{\fetapphi}{\ensuremath{\etapr\phi}}
\newcommand{\etapphi}{\ensuremath{\Bz\ra\fetapphi}}
\newcommand{\Betapphi}{\ensuremath{\calB(\etapphi)}}
\newcommand{\retapphi}{\ensuremath{0.2^{+0.4}_{-0.3}\pm0.1}}

\newcommand{\uletapphi}{\ensuremath{1.0}}

\newcommand{\fetapeppphi}{\ensuremath{\etapr_{\eta\pi\pi} \phi}}
\newcommand{\fetaprgphi}{\ensuremath{\etapr_{\rho\gamma} \phi}}

\newcommand{\fetakz}{\ensuremath{\eta K^0}}

\newcommand{\etakz}{\ensuremath{\Bz\ra\fetakz}}
\newcommand{\retakz}{\ensuremath{1.8^{+0.7}_{-0.6}\pm0.1}}
\newcommand{\Betakz}{\ensuremath{\calB(\etakz)}}

\newcommand{\uletakz}{\ensuremath{2.9}}

\newcommand{\phitoKpKm}{\ensuremath{\phi\ra\Kp\Km}}
\newcommand{\etaprhogam}{\ensuremath{\etapr\ra\rho\gamma}}

\begin{document}

\preprint{\babar-PUB-\BaBarYear/\BaBarNumber} 
\preprint{SLAC-PUB-\SLACPubNumber} 

\begin{flushleft}
\babar-PUB-\BaBarYear/\BaBarNumber \\
 SLAC-PUB-\SLACPubNumber\\
\end{flushleft}

\title{
 \large  \bf\boldmath Searches for \Bz\ Decays to \fetakz, \fetaeta, \fetapetap, \fetaphi, and \fetapphi 
 
}
%
\author{B.~Aubert}
\author{M.~Bona}
\author{D.~Boutigny}
\author{F.~Couderc}
\author{Y.~Karyotakis}
\author{J.~P.~Lees}
\author{V.~Poireau}
\author{V.~Tisserand}
\author{A.~Zghiche}
\affiliation{Laboratoire de Physique des Particules, IN2P3/CNRS et Universit\'e de Savoie,
 F-74941 Annecy-Le-Vieux, France }
\author{E.~Grauges}
\affiliation{Universitat de Barcelona, Facultat de Fisica, Departament ECM, E-08028 Barcelona, Spain }
% ...
\collaboration{The \babar\ Collaboration}
\noaffiliation

\begin{abstract}
We search for $B^0$ meson decays into two-body combinations of $K^0$, $\eta$,
\etapr, and $\phi$ mesons in 324 million \BB\ pairs
 collected with the 
\babar\ detector at the \pep2 asymmetric-energy \epem  collider at SLAC.
We measure the following branching fractions (upper limits at 90\%  confidence level) in units of $10^{-6}$:
 $\Betakz = \retakz (<\uletakz)$,  $\Betaeta = \retaeta (<\uletaeta)$, 
$\Betaphi = \retaphi (<\uletaphi)$, $\Betapphi=\retapphi
(<\uletapphi)$, and  $\Betapetap=\retapetap\ (<\uletapetap)$,
where the first error is statistical and the second systematic.
\end{abstract}


\maketitle

We report the  results of searches for \Bz\ or \Bzb\ meson decays to two charmless
pseudoscalar mesons \cite{charge}  \fetakz, \fetaeta, \fetapetap, and to the
pseudoscalar-vector combinations \fetaphi,
\fetapphi.  None of these decays has  been observed
previously; the published experimental upper limits on their branching
fractions lie in the range $(2-10)\times10^{-6}$ \cite{Isoscalar,etaK0}.
The theoretical predictions for these branching fractions are 
less than a few per million by most estimates
\cite{SU3,CHIANGeta,CHIANGVP,ALI,LEPAGE,BENEKE,SCET}.
Theoretical approaches include those based on flavor SU(3) relations
 \cite{SU3,CHIANGeta,CHIANGVP}, effective Hamiltonians with
factorization and specific $B$-to-light-meson form factors \cite{ALI},
perturbative QCD \cite{LEPAGE}, QCD factorization \cite{BENEKE}, 
and soft collinear effective theory (SCET) \cite{SCET}.
Important advances in the theoretical understanding
    of hadronic charmless two-body B meson decays have
    occurred in the past few years \cite{William}.  With more precise
    experimental results one can test and constrain the
    models.
Improved measurements of decays with
    isoscalar mesons can also help to better understand
    the large difference between the branching fractions for $B\ra \etapr K$ 
and $B \ra \eta K$ decays \cite{William,Lipkin}.

Branching fractions or limits in the $\eta \eta$, $\etapr \etapr$,
$\eta \phi$, and $\etapr \phi$ channels are relevant for the accuracy
with which \CP-violating asymmetry measurements can be interpreted.
The coefficient $S$ of  the \CP-violating sinusoidal factor
in the time evolution of $\etapr K^0$ and $\phi K^0$ can be related to
the CKM phase $\beta = \arg{(-V_{cd}
V^*_{cb}/ V_{td} V^*_{tb})}$ if these decays are dominated by a single
weak phase \cite{PDG2004}.  
Additional higher-order amplitudes with different weak phases would lead
to deviations $\Delta S$ between  the value 
measured in these rare modes and  the precise determination in the
more copious $B^0$ decays to  charmonium-$K^0$ final states.  
SU(3) flavor symmetry \cite{GROSS,Gronau} relates the strength of such
additional amplitudes
    to the decay rates of certain two-body $B^0$ decays, including 
     $\eta \eta$, $\etapr \etapr$, $\eta \phi$, and $\etapr \phi$.

The results presented here are based on data collected
with the \babar\ detector~\cite{BABARNIM}
at the PEP-II asymmetric-energy $e^+e^-$ collider
located at the Stanford Linear Accelerator Center.  An integrated
luminosity of 289~fb$^{-1}$, corresponding to 
$N_{\BB}= 324 $ million \BB\ pairs, was recorded at 
the $\Upsilon (4S)$ resonance (center-of-mass energy $\sqrt{s}=10.58\ \gev$).

Charged particles produced in  \epem\ interactions are detected, and their
momenta measured, by a combination of a vertex tracker, consisting
of five layers of double-sided silicon microstrip detectors, and a
40-layer central drift chamber, both operating in the 1.5 T magnetic
field of a superconducting solenoid. We identify photons and electrons 
using a CsI(Tl) electromagnetic calorimeter.
Further charged-particle identification is provided by the average energy
loss (\dedx ) in the tracking devices and by an internally reflecting
ring-imaging Cherenkov detector (DIRC) covering the central region.

We select $\eta$, $\etapr$, $\phi$, $\rho^0$, $K^0_S$, and $\pi^0$  candidates through the 
decays \etatogg\ (\etagg), \etatoppp\ (\etappp), 
\etaptoepp\ with \etatogg\ (\etapepp), \etaptorg\ (\etaprg), \phitoKpKm,  
$\rho^0\ra\pipi$, $K^0_S \ra \pi^+\pi^-$, and $\pi^0 \ra \gamma\gamma$.  
The photon energy $E_{\gamma}$ must be greater than 30 (100) \mev  for
$\pi^0$ (prompt $\eta$ from $B$) candidates, greater than  200 \mev  in \etaprhogam,
and greater than 50 (100) \mev in \etapepp\ (in the $B \ra \etapepp\ \etapepp$ decay mode).
We make the following requirements on the invariant masses (in \mevcc):
 $490< m_{\gaga}<600$ for \etagg, $120< m_{\gaga}<150$ for $\pi^0$,
 $510<m_{\pi\pi}< 1000$ for $\rho^0$,
$520<m_{\pi\pi\pi}< 570$ for
\etappp, $930<m_{\eta\pi\pi}<990$ for \etapepp,  
$910<m_{\rho\gamma}<1000$ for \etaprg,
 $1005 < m_{K^+K^-} < 1035$ for $\phi$, and $486 < m_{\pi\pi} < 510$
 for $K^0_S$. For $K^0_S$ candidates  we also require    a
vertex $\chi^2$ probability larger than $0.001$ and a reconstructed 
 decay length greater than three times its uncertainty. 
  Secondary charged pions in $\eta$ and \etapr candidates are rejected,
  if their DIRC and 
\dedx\ signatures  are consistent with protons, electrons, or
kaons. Similarly, tracks from $\phi$ decays are
required to be inconsistent with protons, electrons, and pions.
  
A $B$ meson candidate is characterized kinematically by the
energy-substituted 
mass $\mes=\lbrack{(\half s+\pvec_0\cdot\pvec_B)^2/E_0^2-\pvec_B^2}\rbrack^\half$
and energy difference $\DE = E_B^*-\half\sqrt{s}$, where the subscripts $0$ and
$B$ refer to the initial \UfourS\ and to the $B$ candidate, respectively,
and the asterisk denotes the \UfourS\ rest frame. 

Backgrounds arise primarily from random combinations of tracks and
neutral clusters in $\epem\ra\qqbar$ continuum events, where $q = u,
d, s $, or $ c$. 
We reject these events by using the angle
\thetaT\ between the thrust axis of the $B$ candidate in the \UfourS\
frame and that of the rest of the event. 
The  thrust axis
of the $B$ candidate is obtained as the thrust axis of the $B$ decay products.
The distribution of $|\costhr|$ is
sharply peaked near $1.0$ for combinations drawn from jet-like \qqbar\
pairs, and is nearly uniform for $\UfourS \rightarrow \BB$ events.  We require
$|\costhr|<0.9$. To discriminate against $\tau$-pair and two-photon
backgrounds  
we require the event to contain at least  three tracks or one 
track more than the
topology of our final state, whichever is larger.
In decays containing a prompt \etagg\ from $B$ we require  
 $|\mathcal{H_{\eta}}| < 0.9$ 
to remove random combinations with soft photons, where
$\mathcal{H_{\eta}}$ is defined below.
If an event has  multiple $B$ candidates,
 we select the candidate with the highest $B$ vertex $\chi^2$ probability or  using a 
$\chi^2$ quantity computed with the $\eta$ or \etapr\ masses, depending on the decay mode. 
More details on the analysis technique can be found in Ref. \cite{Colo}.

We obtain yields from unbinned extended maximum-likelihood
(ML) fits.  The principal input observables are \DE, \mes, and a
Fisher discriminant \xf\ \cite{Fisher}.
 Where  relevant, the invariant masses \mres\ 
of the intermediate resonances and angular variables \hel\ defined below are used.
The Fisher discriminant \xf\  combines four variables: the angles with respect to 
the beam axis of the $B$ momentum and $B$ thrust axis 
(in the \UfourS\ frame), and the zeroth and second angular moments $L_{0,2}$ 
of the energy flow about the \Bz\ thrust axis.  The moments are defined by
$ L_j = \sum_i p_i\times\left|\cos\theta_i\right|^j$,
where $\theta_i$ is the angle with respect to the $B$ thrust axis of
track or neutral cluster $i$, $p_i$ is its momentum, and the sum
excludes the $B$ candidate.
For $\etagg$ ($\phi$),
$\mathcal{H_{\eta}}$ ($\mathcal{H_{\phi}}$) is defined as the cosine of the angle between 
the direction of a daughter $\gamma$ ($K$) and the flight direction of the
parent of $\eta$ ($\phi$)  in the $\eta$ ($\phi$) rest frame; for \etaprg,
$\mathcal{H_{\rho}}$ is the cosine of the angle between the direction of 
a $\rho$ daughter and the flight direction of the \etapr\ in the
$\rho$ rest frame. The  set of probability density 
functions (PDF) used in 
ML fits, specific to each  decay mode, is determined on the basis of studies
with Monte Carlo (MC) simulated samples \cite{geant}.
We estimate \BB\ backgrounds using MC  samples   of $B$ decays.
The estimated \BB\ background is found to be  negligible for all of
our decay modes except $\eta_{\gamma\gamma} K^0_S$ 
and $\eta_{\gamma\gamma}\phi$.

The extended likelihood function is
\begin{equation}
{\cal L}= \exp{(-\sum_{j=1}^3 n_j)} \prod_{i=1}^N
\left[\sum_{j=1}^3 n_j  {\cal P}_j ({\bf x}_i)\right]\ ,
\end{equation}
where  $N$ is the number of input events,  $n_j$ is the number of
events for hypothesis $j$ ($j=1$ for signal, $j=2$ for continuum
background, and $j=3$ for  \BB\ background), and     ${\cal P}_j ({\bf x}_i)$ is the corresponding PDF evaluated  
with the observables ${\bf x}_i$ of the $i^{th}$ event.  The \BB\
background component is used in the decay modes $\eta_{\gamma\gamma}
K^0_S$ and $\eta_{\gamma\gamma}\phi$.
Since the correlations among the observables in the data are small,
we take each  ${\cal P}_j$  as the product of the PDFs for the separate variables.
We determine the PDF parameters from simulation for the
signal and from sideband
data ($5.25 < \mes\ <5.27$ \gevcc; $0.1<|\DE |<0.2$ \gev ) 
   for continuum  background.  We float some of the continuum  PDF parameters 
in the ML fit. We parameterize each of the functions ${\cal P}_{\rm 1}(\mes),\ 
{\cal  P}_{\rm 1}(\DE),\ { \cal P}_j(\xf),\ $ and the 
peaking components of ${\cal P}_j(\mres)$ with either a Gaussian, the sum of
two Gaussians, or a Crystal Ball function \cite{crystal} as required to describe the 
distribution.  Slowly varying distributions ($ \mres$ and $\DE$  
for combinatorial background, and angular variables) are represented by linear or 
quadratic functions.
The combinatorial background in \mes
is described by the ARGUS function \cite{Argus}.
Large data control samples of $B$ decays to charmed final states of similar 
topology are used to verify the simulated resolutions in \mes and \DE.
Where the control samples reveal differences between data and 
 MC in mass or energy
resolution, we shift or scale the resolution used in the likelihood fits.
The bias in the fit is determined from a large set of simulated experiments, each one with 
the same number of $q \bar{q}$ and signal events as in data. 

\begin{table*}[htp]
\caption{
Fitted signal event  yield, fit bias, detection
efficiency  $\epsilon$, daughter branching fraction product $\prod\calB_i$,
significance $\cal S$, and measured branching
fraction \calB\ with statistical error for each decay mode. For the
combined measurements we give
the significance  (with systematic uncertainties included) and the  branching fraction
with statistical and systematic uncertainty (in parentheses the  90\% CL upper limit).
}
\label{tab:results}
\begin{tabular}{lccccccc}
\dbline
Mode& \quad Yield (ev) \quad&\quad Fit bias (ev) &\quad \eff \quad &\quad
$\prod\calB_i$ (\%) \quad&\quad \signf \quad &\quad \bfemsix \quad  \\
\tbline
~~\fetaggkz  &   $19^{+10}_{-9}$ &$+0.8\pm 0.6$ &$26.7\pm 0.9$&$13.5$&$2.6$ &$1.5^{+0.9}_{-0.8}$ \\
~~\fetapppkz  &  $11^{+6}_{-5}$  &$+1.1\pm 0.4$  &$17.3\pm 0.6$&$7.8$ &$2.7$ &$2.4^{+1.4}_{-1.1}$ \\
\bma{\fetakz} &          &  & &  &\bma{3.5} &  \bma{\retakz}& \bma{~ (< \uletakz)}    \\
\hline
~~\fetaggetagg &   $17^{+10}_{-9}$ &$+3.9\pm0.6$ &$20.8\pm 1.3$&$15.5$&$1.9$&$1.3^{+1.0}_{-0.9}$\\
~~\fetaggetappp &  $10^{+7}_{-5}$  &$+0.5\pm0.4$ &$18.3\pm 1.2$&$17.9$&$2.1$&$0.9^{+0.6}_{-0.5}$\\
~~\fetapppetappp & $2^{+3}_{-2}$   &$+0.3\pm0.4$ &$11.6\pm 0.8$&$5.1$ &$1.1$&$1.1^{+1.6}_{-1.0}$\\
\bma{\fetaeta}&                   &  &  &&\bma{3.0}&\bma{1.1^{+0.5}_{-0.4}\pm 0.1} &\bma{~ (< 1.8)}     \\
\hline
~~\fetaggphi &   $-11^{+7}_{-5}$ &$-2.4\pm0.6$ &$32.3\pm 1.2$&$19.4$&$0.0$&$-0.4^{+0.3}_{-0.2}$ \\
~~\fetapppphi &  $6^{+5}_{-4}$ &$+0.8\pm0.3$ &$20.7\pm 1.0$&$11.1$&$1.5$&$0.7^{+0.7}_{-0.5}$ \\
\bma{\fetaphi} &                   & &&  &\bma{0.0} &\bma{0.1\pm 0.2\pm 0.1}& \bma{~ (<0.6 )}     \\
\hline
~~\fetapeppphi &  $1^{+3}_{-2}$ &$-0.6\pm0.3$  &$23.1\pm 1.1$&$8.6$&$0.8$&$0.3^{+0.5}_{-0.3}$ \\
~~\fetaprgphi &   $-3^{+9}_{-8}$  &$-1.0\pm0.4$&$22.5\pm 0.9$&$14.5$&$0.0$&$-0.2^{+0.9}_{-0.7}$ \\
\bma{\fetapphi} &                   & &&  &\bma{0.5} &\bma{0.2^{+0.4}_{-0.3}\pm 0.1}& \bma{~ (< 1.0)}    \\    
\hline
~~\fetapeppetapepp  &$1^{+2}_{-1}$&$+0.3\pm 0.2$ &$15.2\pm1.0$& $3.1$ &$1.2$&$0.8^{+1.3}_{-0.7}$   \\
~~\fetapeppetaprg   &$9^{+7}_{-5}$&$+1.5\pm 0.3$ &$17.6\pm0.8$& $10.3$&$1.5$&$1.2^{+1.1}_{-0.9}$   \\
\bma{\fetapetap} &                   & &&  &\bma{1.8} &\bma{1.0^{+0.8}_{-0.6}\pm 0.1}& \bma{~ (< 2.4)}    \\
\dbline
\end{tabular}
\vspace{-5mm}
\end{table*}

Table \ref{tab:results} shows the measured yields, efficiencies, 
and products of daughter branching fractions for each decay mode. 
The efficiency is calculated as the ratio of the numbers of signal 
MC events after the cut based selection  to the total generated.  
We compute the branching fractions from the fitted  signal event yields, reconstruction efficiency,
daughter branching fractions, and the number of produced $B$ mesons, assuming equal production 
rates of charged and neutral $B$ pairs at \UfourS. We correct the 
yield for any bias measured with the simulations.
We combine results from different channels by adding the values of
$-2\ln{\cal L}$ (parameterized in terms of the branching fraction), taking into account  the correlated and uncorrelated systematic errors.
We report the statistical significance and the 
branching fractions for the individual decay channels. For the
combined measurements we also report the 90\% confidence level (CL) upper limits.

The statistical error on the signal yield is taken as the change in 
the central value when the quantity $-2\ln{\cal L}$ increases by one 
unit from its minimum value. The significance is taken as the square root 
of the difference between the value of $-2\ln{\cal L}$ (with systematic 
uncertainties included) for zero signal and the value at its minimum.
We determine a Bayesian  90\% CL upper limit assuming a uniform prior
probability distribution by finding the branching 
fraction below which lies 90\% of the total of the likelihood integral 
in the positive branching fraction region.

\begin{figure}[htb]
  \begin{minipage}{\linewidth}
   \begin{center}
    \includegraphics[scale=0.40]{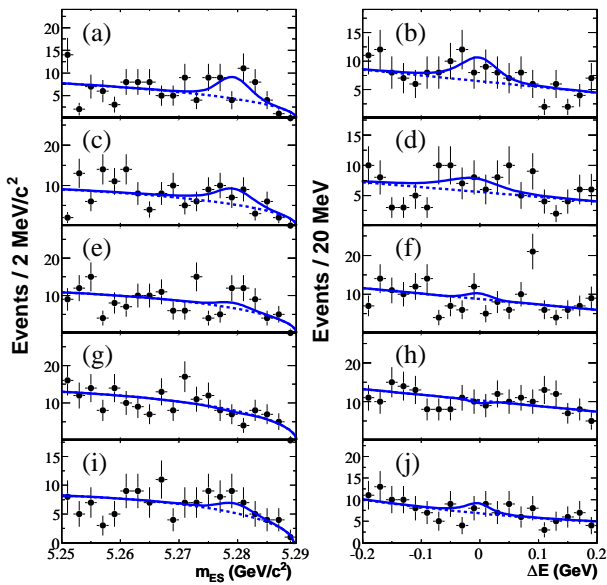}
   \end{center}
   \end{minipage}
   \caption{Signal enhanced  projections on \mes\ (left) and \DE\
     (right) in the decays: (a, b) $\eta \KS$, (c, d)  $\eta \eta$,
     (e, f) $\eta \phi$, (g, h) $\etapr \phi$, (i, j) $\etapr \etapr$.
     Points with error bars (statistical only) represent the data
     (combined measurements), the solid
     line the full fit function, and the dashed line its background
     component.
   }
 \label{fig:projectionDEMes}
 \end{figure}

Figure\ \ref{fig:projectionDEMes} shows, for representative fits, the
 projections onto \mes\ and \DE\ for the five decay modes. The
points show the data after a channel-dependent  requirement on 
the  probability  ratio    
${\cal P}_{\rm{1}}/({\cal P}_{\rm{1}}+{\cal P}_{\rm{2}} +{\cal P}_{\rm{3}}  )$,
 optimized to enhance the
 signal sensitivity and with
the probabilities ${\cal P}_{\rm j}$  evaluated  
without using the variable plotted.  The solid curves show the total
 rescaled fit functions.

The main sources of systematic error include uncertainties in the PDF 
parameterization (0-2 events) and
 ML fit bias (0-2 events).  We evaluate these uncertainties with simulated
experiments by
varying the PDF parameters within their errors and by embedding MC signal events
inside background distributions simulated from PDFs. The uncertainty on $N_{\BB}$
is  1.1\%.  Published world averages
\cite{PDG2004}\ provide the uncertainties in the $B$-daughter branching 
fractions (1-7\%). Other  sources of systematic uncertainty  are
track (1-3\%) and  neutral cluster (2-6\%) reconstruction efficiencies.
The validity of the fit procedure and PDF parameterization, including
the effects of unmodeled correlations among observables, is checked with simulated 
experiments. 

Grossman \etal~\cite{GROSS} introduced a
method to determine a bound on $|\Delta S_f| \equiv |S_f - \stwob|$
where $f$ is a \CP eigenstate produced in charmless $\Bz$ decays
and $S$ is the coefficient of the \CP-violating sinusoidal factor
mentioned above.
The method relies on SU(3) flavor symmetry and
the measured branching fractions of charmless, strangeness-conserving
\Bz decays to constrain the unknown contributions of suppressed
amplitudes in $\Bz\rightarrow f$.
Two of the channels in our study,
$\eta\eta$ and $\etapr\etapr$,
are relevant to the $\Delta S_f$ bound for $f=\etapr\Kz$, while two others,
$\eta\phi$ and $\etapr\phi$,
are relevant for $f=\phi\Kz$.
Using the technique described in Ref.~\cite{Gary} and  evaluating
90\% CL upper limits, 
we find $|\Delta S_{\etapr\Kz}|<0.15$ and $|\Delta S_{\phi\Kz}|<0.38$.
This  new $\Delta S_{\etapr\Kz}$ bound also makes use of
our recent results~\cite{EtaprEta2006} on the
$\Bz\rightarrow\etapr\eta$, $\etapr\piz$, and $\eta\piz$ channels. 

In summary, we present updated measurements 
of branching fractions for five \Bz\ decays to charmless  meson
pairs. Our results represent substantial improvements on the previous upper
limits \cite{Isoscalar,etaK0}.

We are grateful for the excellent luminosity and machine conditions
provided by our \pep2\ colleagues, 
and for the substantial dedicated effort from
the computing organizations that support \babar.
The collaborating institutions wish to thank 
SLAC for its support and kind hospitality. 
This work is supported by
DOE
and NSF (USA),
NSERC (Canada),
IHEP (China),
CEA and
CNRS-IN2P3
(France),
BMBF and DFG
(Germany),
INFN (Italy),
FOM (The Netherlands),
NFR (Norway),
MIST (Russia), and
PPARC (United Kingdom). 
Individuals have received support from the
Marie Curie EIF (European Union) and
the A.~P.~Sloan Foundation.


\begin{thebibliography}{99}
\bibitem{charge}
The charge conjugate of the named state is implicitly included, here
and throughout this paper.
\end{thebibliography}
\end{document}